\documentclass[american,aps,prl,reprint]{revtex4-1}
\usepackage[T1]{fontenc}
\usepackage[latin9]{inputenc}
\usepackage{amsmath}
\usepackage{amssymb}
\usepackage{graphicx}

\makeatletter

\providecommand{\tabularnewline}{\\}

\@ifundefined{textcolor}{}
{%
 \definecolor{BLACK}{gray}{0}
 \definecolor{WHITE}{gray}{1}
 \definecolor{RED}{rgb}{1,0,0}
 \definecolor{GREEN}{rgb}{0,1,0}
 \definecolor{BLUE}{rgb}{0,0,1}
 \definecolor{CYAN}{cmyk}{1,0,0,0}
 \definecolor{MAGENTA}{cmyk}{0,1,0,0}
 \definecolor{YELLOW}{cmyk}{0,0,1,0}
 }

\usepackage{times}
\usepackage{txfonts}
\usepackage{braket}

\makeatother

\usepackage{babel}
\begin{document}

\title{Limits for entanglement distribution with separable states}

\author{Alexander Streltsov}

\email{streltsov@thphy.uni-duesseldorf.de}

\author{Hermann Kampermann}

\author{Dagmar Bruß}

\affiliation{Heinrich-Heine-Universität Düsseldorf, Institut für Theoretische
Physik III, D-40225 Düsseldorf, Germany}
\begin{abstract}
Entanglement distribution with separable states has recently attracted
considerable attention. Recent results suggest that quantum discord
-- a measure for quantum correlations beyond entanglement -- is responsible
for this counterintuitive phenomenon. In this work we study this question
from a different perspective, and find minimal requirements for a
separable state to be useful for entanglement distribution. Surprisingly,
we find that the presence of quantum discord is not sufficient to
ensure entanglement distribution: there exist states with nonzero
quantum discord which nevertheless cannot be used for entanglement
distribution. As a result, we show that entanglement distribution
is not possible with rank two separable states. Our work sheds new
light on the task of entanglement distribution with separable states,
and reveals a new classification of quantum states with respect to
their usefulness for this task.
\end{abstract}
\maketitle
A fundamental task in quantum information processing is the distribution
of entanglement between two distant parties. It has been shown in
\cite{Cubitt2003} that, counterintuitively, this task can be achieved
by sending a particle which exhibits no entanglement with the rest
of the system. Very recently, such entanglement distribution with
separable states has been demonstrated experimentally in physical
systems with continuous \cite{Vollmer,Peuntinger} and discrete variables
\cite{Fedrizzi}. Quantum discord, a novel type of quantum correlations
going beyond entanglement \cite{Zurek2000,Ollivier2001,Henderson2001},
has been identified as the figure of merit for this puzzling phenomenon
\cite{Streltsov2012,Chuan2012}. This finding is in accordance with
earlier results, supporting the crucial role of quantum discord and
related quantifiers of quantum correlations \cite{Luo2008,Modi2010,Giorda2010,Adesso2010,Dakic2010,Giorgi2011}
in quantum information theory. These results include thermodynamical
approaches \cite{Oppenheim2002,Zurek2003}, and the relations to entanglement
creation in the quantum measurement process \cite{Streltsov2011,Piani2011,Adesso}
and to entanglement consumption in quantum state merging \cite{Madhok2011,Cavalcanti2011}.
Recently, the role of quantum discord for quantum metrology \cite{Modi2011,Girolami2013},
encoding \cite{Gu2012} and sharing \cite{Zwolak2013,Streltsov2013}
of information has also been subjected to scrutiny. Quantum discord
was further proposed to be the figure of merit for the quantum computing
protocol known as DQC1 \cite{Datta2008} and for the task of remote
state preparation \cite{Dakic2012}. While some of the arguments are
still controversial \cite{Dakic2010,Giorgi2013,Horodecki}, they have
led to a fruitful debate about the role of general quantum correlations
in quantum information theory which goes on until the present day
\cite{Merali2011,Modi2012}.

\begin{figure}
\begin{centering}
\includegraphics[width=0.7\columnwidth]{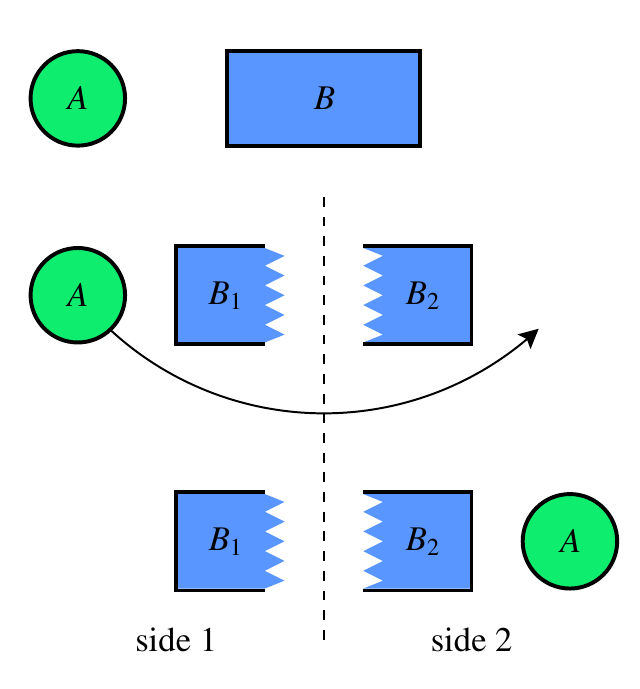} 
\par\end{centering}

\caption{\label{fig:1} A quantum system consisting of two parties $A$ and
$B$ (upper figure) is useful for entanglement distribution, if the
party $B$ can be divided in two parties $B_{1}$ and $B_{2}$ (middle
figure) in such a way, that sending the particle $A$ from side $1$
to side $2$ leads to an increase of entanglement (lower figure).}
\end{figure}

In this Letter, we aim to find minimal requirements for entanglement
distribution with separable states. To this end we consider a general
distribution protocol, and identify properties for a separable state
to be a resource for entanglement distribution. In the following,
we call a (not necessarily separable) quantum state $\rho^{AB}$ \emph{useful}
for entanglement distribution, if it is possible to divide the party
$B$ in two parties $B_{1}$ and $B_{2}$ in such a way, that sending
the particle $A$ from side 1 to side 2 leads to an increase of entanglement
\footnote{Here, ``dividing'' $B$ into $B_{1}$ and $B_{2}$ means to define
a tensor product structure for the corresponding Hilbert spaces: ${\cal H}_{B}={\cal H}_{B_{1}}\otimes{\cal H}_{B_{2}}$.%
}: 
\begin{equation}
E^{B_{1}|B_{2}A}>E^{AB_{1}|B_{2}},
\end{equation}
see also Fig. \ref{fig:1}. Certainly, any entangled state is useful
for entanglement distribution, as can be seen by giving the full system
$B$ to side $1$, i.e., $B_{1}=B$. This implies that the presence
of entanglement between $A$ and $B$ is sufficient for entanglement
distribution. On the other hand, the finding that entanglement can
be distributed with separable states \cite{Cubitt2003} demonstrates
that the presence of entanglement between $A$ and $B$ is not necessary,
and that, in general, some other kind of quantum correlations beyond
entanglement is responsible for this process.

Recently, quantum discord was identified as the key resource for entanglement
distribution: the distribution of any finite amount of entanglement
needs the transmission of at least the same amount of quantum discord
\cite{Streltsov2012,Chuan2012}. These results show that -- in contrast
to entanglement -- quantum discord is implicitly required, if two
parties wish to increase the amount of entanglement between them.
As a consequence, all classical-quantum states, i.e., states of the
form 
\begin{equation}
\rho_{\mathrm{cq}}=\sum_{i}p_{i}\ket{i}\bra{i}^{A}\otimes\rho_{i}^{B}
\end{equation}
 with $\braket{i|j}=\delta_{ij}$, cannot be used for entanglement
distribution, since all those states have zero quantum discord \cite{Ollivier2001}.
On the other hand, the results presented in \cite{Streltsov2012,Chuan2012}
support the intuition that the presence of quantum discord in a quantum
state $\rho^{AB}$ already ensures its usefulness for entanglement
distribution. This idea leads us to the main question of this Letter:
\emph{Are all states with nonzero quantum discord useful for entanglement
distribution?}

For approaching the answer to this question, we first consider the
most simple class of potentially useful separable states: 
\begin{equation}
\rho^{AB}=p\ket{\psi_{1}}\bra{\psi_{1}}^{A}\otimes\ket{\phi_{1}}\bra{\phi_{1}}^{B}+(1-p)\ket{\psi_{2}}\bra{\psi_{2}}^{A}\otimes\ket{\phi_{2}}\bra{\phi_{2}}^{B},\label{eq:rho}
\end{equation}
which is a mixture of two product states $\ket{\psi_{1}^{A}}\ket{\phi_{1}^{B}}$
and $\ket{\psi_{2}^{A}}\ket{\phi_{2}^{B}}$ with corresponding probabilities
$p>0$ and $(1-p)>0$. Noting that this state has nonzero discord
for a generic choice of the states $\ket{\psi_{i}^{A}}$, $\ket{\phi_{i}^{B}}$
and the probability $p$, it is reasonable to conjecture that this
state is generically useful for entanglement distribution. Surprisingly,
as we will see in the following, this intuition is not correct. This
implies that the answer to the question stated above is negative,
leading to strong limitations on entanglement distribution with separable
states.

In the following we will show that the state given in Eq. (\ref{eq:rho})
cannot be used for entanglement distribution, regardless of the choice
of the states $\ket{\psi_{i}^{A}}$, $\ket{\phi_{i}^{B}}$ and the
probability $p$. In particular, we will show that for any division
of the party $B$ in two parties $B_{1}$ and $B_{2}$, sending the
particle $A$ from one side to the other will never change the amount
of entanglement: 
\begin{equation}
E_{n}^{B_{1}|B_{2}A}=E_{n}^{AB_{1}|B_{2}}.\label{eq:En}
\end{equation}
Here, $E_{n}$ is the \emph{negativity}, defined for a state $\rho=\rho^{XY}$
as $E_{n}^{X|Y}=\sum_{i}\left|\lambda_{i}\right|$, where $\lambda_{i}<0$
are the negative eigenvalues of the matrix $\rho^{T_{X}}$ \cite{.Zyczkowski1998,Vidal2002},
and the superscript $T_{X}$ denotes partial transposition with respect
to the party or parties in $X$. As the only computable quantifier
of entanglement known today, the negativity is widely used in quantum
information theory \cite{Horodecki2009}. Since the negativity is
zero only on separable and bound entangled states \cite{Horodecki1998},
and bound entanglement is known to be absent in bipartite states with
rank smaller than four \cite{Horodecki2000,Horodecki2003,Chen2008},
$E_{n}$ is a faithful quantifier of entanglement for the states presented
in Eq. (\ref{eq:rho}).

For proving Eq. (\ref{eq:En}) we consider the partially transposed
density matrices $\rho^{T_{B_{1}}}$ and $\rho^{T_{AB_{1}}}$ of the
total state $\rho=\rho^{AB}=\rho^{AB_{1}B_{2}}$ given in Eq. (\ref{eq:rho}),
where $B_{1}$ and $B_{2}$ are two subsystems of $B$. In particular,
we will show that the matrices $\rho^{T_{B_{1}}}$ and $\rho^{T_{AB_{1}}}$
are equal up to a unitary, and thus share the same set of eigenvalues.
For showing this, we start with the partially transposed density matrix
\begin{equation}
\rho^{T_{B_{1}}}=p\ket{\psi_{1}}\bra{\psi_{1}}^{A}\otimes M_{1}^{B}+(1-p)\ket{\psi_{2}}\bra{\psi_{2}}^{A}\otimes M_{2}^{B}
\end{equation}
with $M_{1}^{B}=(\ket{\phi_{1}}\bra{\phi_{1}})^{T_{B_{1}}}$ and $M_{2}^{B}=(\ket{\phi_{2}}\bra{\phi_{2}})^{T_{B_{1}}}$.
In the next step we will show that a partial transposition of this
matrix $\rho^{T_{B_{1}}}$ with respect to the subsystem $A$ is equivalent
to a unitary, i.e., $\rho^{T_{AB_{1}}}=U\rho^{T_{B_{1}}}U^{\dagger}$.
This can be seen by considering the Bloch vectors $\mathbf{r}$ and
$\mathbf{s}$ corresponding to the states $\ket{\psi_{1}^{A}}$ and
$\ket{\psi_{2}^{A}}$, i.e., $\ket{\psi_{1}}\bra{\psi_{1}}^{A}=\frac{1}{2}(\openone+\sum_{i}r_{i}\sigma_{i})$
and $\ket{\psi_{2}}\bra{\psi_{2}}^{A}=\frac{1}{2}(\openone+\sum_{i}s_{i}\sigma_{i})$
with Pauli matrices $\sigma_{i}$ %
\footnote{Since $\ket{\psi_{1}^{A}}$ and $\ket{\psi_{2}^{A}}$ span a two-dimensional
subspace, this argument is not limited to qubits, but applies regardless
of the dimension of the party $A$.%
}. The transposition of the states $\ket{\psi_{1}^{A}}$ and $\ket{\psi_{2}^{A}}$
takes them to new states $\ket{\tilde{\psi}_{1}^{A}}$ and $\ket{\tilde{\psi}_{2}^{A}}$
with Bloch vectors $\tilde{\mathbf{r}}$ and $\tilde{\mathbf{s}}$.
At this point, it is crucial to note that the product of the Bloch
vectors does not change under transposition: $\tilde{\mathbf{r}}\cdot\tilde{\mathbf{s}}=\mathbf{r}\cdot\mathbf{s}$.
This implies that the transposition of the subsystem $A$ is equivalent
to a joint rotation of the Bloch vectors $\mathbf{r}\rightarrow\tilde{\mathbf{r}}$
and $\mathbf{s}\rightarrow\tilde{\mathbf{s}}$, which on the other
hand corresponds to a unitary acting on the subsystem $A$. This proves
that the matrices $\rho^{T_{B_{1}}}$ and $\rho^{T_{AB_{1}}}$ are
equal up to a unitary, implying that the eigenvalues of both matrices
must be the same. Starting from this result, Eq. (\ref{eq:En}) is
seen to be correct by recalling that the negativity $E_{n}^{X|Y}$
depends only on the eigenvalues of the partially transposed density
matrix $\rho^{T_{X}}$.

\begin{figure}[b]
\begin{centering}
\includegraphics[width=1\columnwidth]{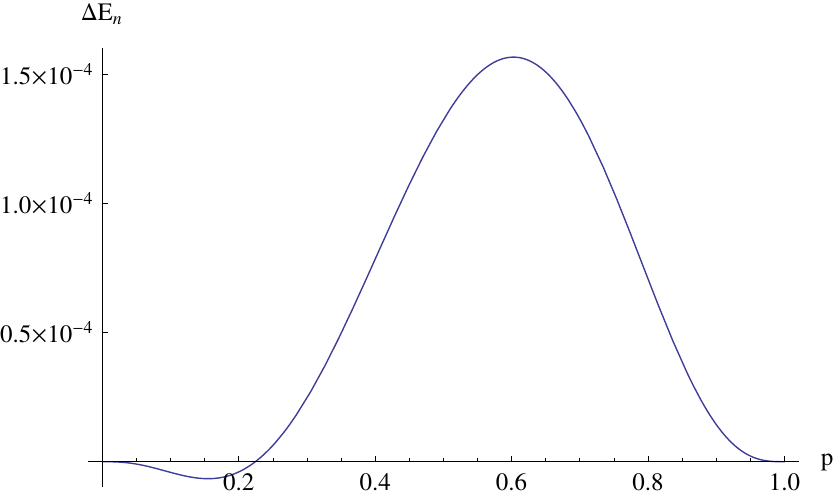} 
\par\end{centering}

\centering{}\caption{\label{fig:2} Entanglement distribution with separable states: the
plot shows the amount of distributed entanglement $\Delta E_{n}=E_{n}^{B_{1}|B_{2}A}-E_{n}^{AB_{1}|B_{2}}$
for the state given in Eq. (\ref{eq:rho2}) as function of the probability
$p$. The state allows to distribute a finite amount of entanglement
$\Delta E_{n}>0$ in the range $p'<p<1$ with $p'\approx0.22$.}
\end{figure}
The results presented so far imply crucial constraints on the possibility
to distribute entanglement with separable states. In particular we
have seen that the distribution of entanglement is not possible, if
the corresponding separable state is a mixture of two pure product
states, see Eq. (\ref{eq:rho}). In the next step we will see that
this limitation can be surpassed, if the pure states $\ket{\psi_{i}^{A}}$
in Eq. (\ref{eq:rho}) are replaced by mixed states $\rho_{i}^{A}$.
In this case the total state takes the form 
\begin{equation}
\rho^{AB}=p\cdot\rho_{1}^{A}\otimes\ket{\phi_{1}}\bra{\phi_{1}}^{B}+(1-p)\cdot\rho_{2}^{A}\otimes\ket{\phi_{2}}\bra{\phi_{2}}^{B}.\label{eq:rho2}
\end{equation}
The use of this state for entanglement distribution can be demonstrated
by a proper choice of the states $\rho_{i}^{A}$ and $\ket{\phi_{i}^{B}}$.
This can be achieved by defining the states $\rho_{i}^{A}$ of the
subsystem $A$ as follows: \begin{subequations}\label{eq:rho12}

\begin{eqnarray}
\rho_{1}^{A} & = & \frac{1}{4}\ket{0}\bra{0}+\frac{3}{4}\ket{1}\bra{1},\\
\rho_{2}^{A} & = & \frac{1}{2}\ket{a}\bra{a}+\frac{1}{2}\ket{b}\bra{b},
\end{eqnarray}
\end{subequations} where $\ket{a}$ and $\ket{b}$ are nonorthogonal
\emph{qutrit} \emph{states}, defined as $\ket{a}=(\ket{0}+\ket{1}+\ket{2})/\sqrt{3}$
and $\ket{b}=(\ket{0}+i\ket{1})/\sqrt{2}$. Finally, the party $B$
consists of two subsystems $B_{1}$ and $B_{2}$, and the corresponding
states $\ket{\phi_{i}^{B}}$ can be chosen as $\ket{\phi_{1}^{B}}=(\ket{00}+\ket{01}+i\ket{11})/\sqrt{3}$
and $\ket{\phi_{2}^{B}}=(\sqrt{8}\ket{00}+\ket{11})/3$. As can be
seen from the difference $\Delta E_{n}=E_{n}^{B_{1}|B_{2}A}-E_{n}^{AB_{1}|B_{2}}$,
shown in Fig. \ref{fig:2} as a function of $p$, this particular
setting allows to distribute a finite amount of entanglement $\Delta E_{n}>0$
in the range $p'<p<1$ for $p'\approx0.22$. 

The example presented above should be regarded as a proof of principle:
it explicitly demonstrates that some separable states which are mixtures
of only two product states can -- in principle -- be used for entanglement
distribution. In particular, we have seen that a successful distribution
of entanglement can be achieved by a specific choice of mixed\emph{
qutrit states} $\rho_{1}^{A}$ and $\rho_{2}^{A}$, see Eq. (\ref{eq:rho12}).
As we will see in the following, it is indeed crucial that the transmitted
particle $A$ is not a qubit: for entanglement distribution with separable
states as given in Eq. (\ref{eq:rho2}) the dimension of $A$ needs
to be at least three. For proving this statement it is enough to show
that for a two-dimensional subsystem $A$ the state given in Eq. (\ref{eq:rho2})
cannot be used for entanglement distribution, i.e., Eq. (\ref{eq:En})
is satisfied. This can be seen by observing that the arguments given
in the proof of Eq. (\ref{eq:En}) for mixtures of two \emph{pure}
product states remain valid if the pure states $\ket{\psi_{1}^{A}}$
and $\ket{\psi_{2}^{A}}$ are replaced by arbitrary qubit states $\rho_{1}^{A}$
and $\rho_{2}^{A}$.

On the one hand, we have seen that entanglement distribution with
separable states is impossible, if the separable state is a mixture
of two pure product states only. On the other hand, we have also demonstrated
a possibility to avoid this problem by using two mixed states $\rho_{1}^{A}$
and $\rho_{2}^{A}$ for the exchanged particle $A$. In the next step
we will show that mixedness of \emph{both} states is essential: entanglement
distribution is not possible if $\rho_{1}^{A}$ or $\rho_{2}^{A}$
is pure, regardless of the dimension of the exchanged particle $A$.
We will prove this statement by showing that Eq. (\ref{eq:En}) is
satisfied for all states given in Eq. (\ref{eq:rho2}) as long as
either $\rho_{1}^{A}$ or $\rho_{2}^{A}$ is pure. Without loss of
generality we can assume that $\rho_{1}^{A}=\ket{\psi}\bra{\psi}^{A}$
is pure, and the state $\rho_{2}^{A}=\tau^{A}$ is diagonal in the
computational basis: $\tau^{A}=\sum_{i}\lambda_{i}\ket{i}\bra{i}^{A}$
\footnote{If it was possible to violate Eq. (\ref{eq:En}) by choosing $\tau^{A}$
which is not diagonal in the computational basis, the invariance of
negativity under local unitaries implies that Eq. (\ref{eq:En}) is
also violated for the diagonal state $\tau{}_{\mathrm{diag}}^{A}=U\tau^{A}U^{\dagger}$.%
}. Using similar lines of reasoning as above we will prove the validity
of Eq. (\ref{eq:En}) by showing that the partially transposed density
matrices $\rho^{T_{B_{1}}}$ and $\rho^{T_{AB_{1}}}$ are equal up
to a unitary. In particular, the matrix $\rho^{T_{B_{1}}}$ has now
the form 
\begin{equation}
\rho^{T_{B_{1}}}=p\ket{\psi}\bra{\psi}^{A}\otimes M_{1}^{B}+(1-p)\cdot\tau^{A}\otimes M_{2}^{B},
\end{equation}
where $M_{1}^{B}=(\ket{\phi_{1}}\bra{\phi_{1}})^{T_{B_{1}}}$ and
$M_{2}^{B}=(\ket{\phi_{2}}\bra{\phi_{2}})^{T_{B_{1}}}$. The matrix
$\rho^{T_{AB_{1}}}$ can be obtained from this expression by performing
partial transposition on the subsystem $A$: 
\begin{equation}
\rho^{T_{AB_{1}}}=p\ket{\tilde{\psi}}\bra{\tilde{\psi}}^{A}\otimes M_{1}^{B}+(1-p)\cdot\tau^{A}\otimes M_{2}^{B}.
\end{equation}
Here, we used the fact that $\tau^{A}=\sum_{i}\lambda_{i}\ket{i}\bra{i}^{A}$
is diagonal in the computational basis, and thus does not change under
transposition. The relation between the state $\ket{\psi^{A}}$ and
the transposed state $\ket{\tilde{\psi}^{A}}$ can be seen by expanding
both states in the computational basis: $\ket{\tilde{\psi}^{A}}=\sum_{j}c_{j}^{*}\ket{j^{A}}$,
where $c_{j}$ are the coefficients of the state $\ket{\psi^{A}}$,
i.e., $\ket{\psi^{A}}=\sum_{j}c_{j}\ket{j^{A}}$. In the final step
it is important to note that the states $\ket{\psi^{A}}$ and $\ket{\tilde{\psi}^{A}}=U\ket{\psi^{A}}$
are related by the unitary $U=\sum_{j}e^{-2i\phi_{j}}\ket{j}\bra{j}^{A}$,
where $\phi_{j}$ is the phase corresponding to the coefficient $c_{j}=|c_{j}|\cdot e^{i\phi_{j}}$.
Since this unitary is diagonal in the computational basis, it does
not change the state $\tau^{A}$, and thus we obtain the desired result:
$\rho^{T_{AB_{1}}}=U\rho^{T_{B_{1}}}U^{\dagger}$. This proves that
for a successful distribution of entanglement with separable states
as given in Eq. (\ref{eq:rho2}) both states $\rho_{1}^{A}$ and $\rho_{2}^{A}$
must be mixed.

The results presented above indicate that the structure of the separable
state -- and in particular the number of product states in its mixture
-- is crucial, if the separable state is to be used as a resource
for entanglement distribution. While a mixture of only two product
states does not allow to distribute any entanglement by sending a
single qubit, this limitation disappears if the exchanged particle
has dimension three. We also note that this result remains valid if
the pure states $\ket{\phi_{i}^{B}}$ of the subsystem $B$ are replaced
by mixed states $\rho_{i}^{B}$. In particular, separable states of
the form $\rho^{AB}=p\cdot\rho_{1}^{A}\otimes\rho_{1}^{B}+(1-p)\cdot\rho_{2}^{A}\otimes\rho_{2}^{B}$
can only be used for entanglement distribution if the transmitted
particle $A$ has at least dimension three, and if both states $\rho_{1}^{A}$
and $\rho_{2}^{A}$ are not pure.

\begin{figure}[b]
\begin{centering}
\includegraphics[width=1\columnwidth]{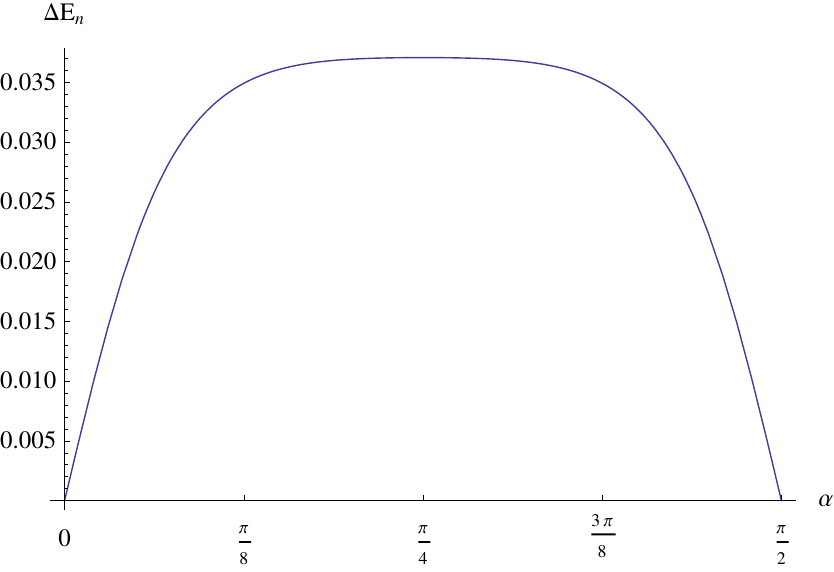} 
\par\end{centering}

\caption{\label{fig:3}Entanglement distribution with separable states by sending
a single qubit: the plot shows the amount of distributed entanglement
$\Delta E_{n}=E_{n}^{B_{1}|B_{2}A}-E_{n}^{AB_{1}|B_{2}}$ for the
state given in Eq. (\ref{eq:rho3}) as function of the parameter $\alpha$.
A finite amount of entanglement $\Delta E_{n}>0$ can be distributed
in the range $0<\alpha<\pi/2$.}
\end{figure}
 In the next step it is natural to ask about the situation where the
separable state used for entanglement distribution is more general.
As we will see in the following, qubits can still be used to distribute
entanglement if the separable state is a mixture of at least \emph{three}
product states. This can be demonstrated on the following separable
state: 
\begin{equation}
\rho^{AB}=\frac{1}{3}\sum_{i=1}^{3}\ket{\psi_{i}}\bra{\psi_{i}}^{A}\otimes\ket{\phi_{i}}\bra{\phi_{i}}^{B},\label{eq:rho3}
\end{equation}
where the \emph{qubit states} $\ket{\psi_{i}^{A}}$ of the transmitted
particle $A$ are chosen as follows: $\ket{\psi_{1}^{A}}=(\ket{0}+\ket{1})/\sqrt{2}$,
$\ket{\psi_{2}^{A}}=(\ket{0}+i\ket{1})/\sqrt{2}$, and $\ket{\psi_{3}^{A}}=\ket{0}$.
The party $B$ consists of two subsystems $B_{1}$ and $B_{2}$, and
the corresponding states $\ket{\phi_{i}^{B}}$ are defined as $\ket{\phi_{1}^{B}}=\ket{01}$,
$\ket{\phi_{2}^{B}}=(\ket{00}+i\ket{11})/\sqrt{2}$, and $\ket{\phi_{3}^{B}}=\cos\alpha\ket{00}+\sin\alpha\ket{11}$.
As can be seen from Fig. \ref{fig:3}, this state allows to distribute
a finite amount of entanglement $\Delta E_{n}=E_{n}^{B_{1}|B_{2}A}-E_{n}^{AB_{1}|B_{2}}$
in the range $0<\alpha<\pi/2$, where $\alpha$ is the parameter of
the state $\ket{\phi_{3}^{B}}$. 

As will become clear in a moment, the reason why the state in Eq.
(\ref{eq:rho3}) is useful for entanglement distribution lies in the
structure of the states $\ket{\psi_{i}^{A}}$. In particular, their
Bloch vectors are given by $\mathbf{r}_{1}=(1,0,0)^{T}$, $\mathbf{r}_{2}=(0,1,0)^{T}$,
and $\mathbf{r}_{3}=(0,0,1)^{T}$. Observe that these three vectors
are linearly independent, i.e., they are not all in the same plane.
We will see in the following that this feature is crucial for entanglement
distribution, where at the same time we will generalize our results
to arbitrary separable states, i.e., states of the form 
\begin{equation}
\rho^{AB}=\sum_{i}p_{i}\cdot\rho_{i}^{A}\otimes\rho_{i}^{B},
\end{equation}
where the exchanged particle $A$ is a qubit. Following the same arguments
as in the preceding discussion we consider the partially transposed
matrices \begin{subequations} 
\begin{eqnarray}
\rho^{T_{B_{1}}} & = & \sum_{i}p_{i}\cdot\rho_{i}^{A}\otimes M_{i}^{B},\\
\rho^{T_{AB_{1}}} & = & \sum_{i}p_{i}\cdot\tilde{\rho}_{i}^{A}\otimes M_{i}^{B}
\end{eqnarray}
\end{subequations} with $M_{i}^{B}=(\rho_{i}^{B})^{T_{B_{1}}}$.
Recall that the state $\rho^{AB}$ cannot be used for entanglement
distribution if the two matrices $\rho^{T_{B_{1}}}$ and $\rho^{T_{AB_{1}}}$
are equal up to a unitary. On the one hand, this is the case whenever
the transposition on the subsystem $A$ of the matrix $\rho^{T_{B_{1}}}$
corresponds to a joint rotation of the Bloch vectors $\mathbf{r}_{i}\rightarrow\tilde{\mathbf{r}}_{i}$,
i.e., whenever there exists a special orthogonal $3\times3$ matrix
$O$ such that $\tilde{\mathbf{r}}_{i}=O\cdot\mathbf{r}_{i}$ %
\footnote{Note that this argument applies only if the exchanged particle $A$
is a qubit, since in this case any rotation on the Bloch sphere corresponds
to a unitary on the corresponding Hilbert space. The fact that this
is no longer true for higher-dimensional systems is responsible for
the effect presented in Eq. (\ref{eq:rho2}) and Fig. \ref{fig:2}:
it allows to distribute entanglement with mixtures of only two product
states, if the exchanged particle is a qutrit.%
}. Here, $\mathbf{r}_{i}$ and $\tilde{\mathbf{r}}_{i}$ are the Bloch
vectors of $\rho_{i}^{A}$ and the transposed state $\tilde{\rho}_{i}^{A}$,
respectively. On the other hand, it is crucial to note that the Bloch
vector $\tilde{\mathbf{r}}$ corresponding to a transposed state $\tilde{\rho}=\rho^{T}$
is related to the Bloch vector $\mathbf{r}$ of the initial state
$\rho$ via a reflection on the xz-plane, i.e., $(\tilde{r}_{1},\tilde{r}_{2},\tilde{r}_{3})=(r_{1},-r_{2},r_{3})$.
Combining these results we can say that $\rho^{AB}$ cannot be used
for entanglement distribution if for all the Bloch vectors $\mathbf{r}_{i}$
a reflection on the xz-plane is equivalent to a rotation. Note that
this is always fulfilled if the number of Bloch vectors is two, in
accordance with the finding that mixtures of two product states cannot
be used for entanglement distribution by sending a qubit. For more
than two vectors a reflection does not necessarily correspond to a
rotation, supporting the finding that qubits can be used for entanglement
distribution if the number of product states is more than two. Finally,
we point out that reflection is equivalent to rotation for any number
of Bloch vectors, whenever all the Bloch vectors are in the same plane,
i.e., whenever any Bloch vector $\mathbf{r}_{i}$ can be written as
a superposition of $\mathbf{r}_{1}$ and $\mathbf{r}_{2}$. This immediately
leads to a generalization of our previous results: entanglement distribution
with separable states by sending a single qubit is only possible if
the corresponding Bloch vectors $\mathbf{r}_{i}$ are not all in the
same plane.

In conclusion, we established minimal requirements for a separable
state to be a resource for entanglement distribution. Here, both the
dimension of the exchanged particle and the number of product terms
in the decomposition play a crucial role -- for a summary see Table
\ref{table:1}. These requirements were deduced from general symmetry
arguments, relating the partial transpose to a rotation. Our results
provide an answer to the main question of this Letter: there are states
with nonzero quantum discord which cannot be used as a resource for
entanglement distribution. In particular, we have shown that a separable
state cannot be used for this task, if it is a mixture of only two
pure product states. Since all rank two separable states are mixtures
of two pure product states \cite{Sanpera1998}, we can conclude that
entanglement distribution with separable states requires states with
rank at least three. Finally, our results suggest a new classification
of quantum states according to their usefulness for entanglement distribution.
While all entangled states are evidently useful for this task, a separable
state can only be useful if it exhibits nonzero quantum discord, and
fulfills the additional requirements provided in Table \ref{table:1}.
The induced substructure within the set of separable states is reminiscent
of the structure within the set of entangled states, which arises
from being useful for distillation (free entanglement) or not being
useful for distillation (bound entanglement). Due to the fact that
entanglement distribution is a fundamental task in quantum information
theory, the results presented in this work may be helpful to devise
new protocols for entanglement distribution, and to gain new insights
into the properties of quantum entanglement and general quantum correlations.

\emph{Acknowledgements}: We acknowledge financial support by the German
Federal Ministry of Education and Research (BMBF, project QuOReP),
and ELES.
\begin{table}
\begin{centering}
\begin{ruledtabular}%
\begin{tabular}{ccc}
$d_{A}$ & $n$ & Entanglement distribution possible?\tabularnewline
\hline 
$2$ & $2$ & no\tabularnewline
$2$ & $3$ & yes\tabularnewline
$3$ & $2$ & yes\tabularnewline
\end{tabular}\end{ruledtabular}
\par\end{centering}

\caption{\label{table:1} Requirements on a separable state $\rho^{AB}=\sum_{i=1}^{n}p_{i}\cdot\rho_{i}^{A}\otimes\rho_{i}^{B}$
to be useful for entanglement distribution. Here, $d_{A}$ is the
dimension of the exchanged particle $A$, and $n$ is the minimal
number of product terms in the decomposition.}
\end{table}

\bibliographystyle{apsrev4-1}
\bibliography{literature}

\end{document}